\documentclass[sigconf]{acmart}

\AtBeginDocument{%
  \providecommand\BibTeX{{%
    \normalfont B\kern-0.5em{\scshape i\kern-0.25em b}\kern-0.8em\TeX}}}

\copyrightyear{2024}
\acmYear{2024}
\setcopyright{acmlicensed}
\acmConference[UbiComp Companion '24]{Companion of the 2024 ACM International Joint Conference on Pervasive and Ubiquitous Computing }{October 5--9, 2024}{Melbourne, VIC, Australia}
\acmBooktitle{Companion of the 2024 ACM International Joint Conference on Pervasive and Ubiquitous Computing (UbiComp Companion '24), October 5--9, 2024, Melbourne, VIC, Australia}
\acmDOI{10.1145/3675094.3678494}
\acmISBN{979-8-4007-1058-2/24/10}

\usepackage{multirow}

\usepackage{tabularx}

\begin{document}


\title[Exploring LLMs to Evaluate EEG-Based Multimodal Data]{Exploring Large-Scale Language Models to Evaluate EEG-Based Multimodal Data for Mental Health}

\author{Yongquan Hu}
\orcid{0000-0003-1315-8969}
 \email{yongquan.hu@unsw.edu.au}
\affiliation{%
 \institution{University of New South Wales}
 \streetaddress{Kingsford}
 \city{Sydney}
 \state{NSW}
 \country{Australia}}

\author{Shuning Zhang}
\orcid{0000-0002-4145-117X}
\email{zsn23@mails.tsinghua.edu.cn}
\affiliation{%
  \institution{Tsinghua University}
  \city{Beijing}
  \country{China}
}

\author{Ting Dang}
\orcid{0000-0003-3806-1493}
\email{ting.dang@unimelb.edu.au}
\affiliation{%
  \institution{University of Melbourne}
  \city{Melbourne}
  \state{VIC}
  \country{Australia}
}

\author{Hong Jia}
\orcid{0000-0002-6047-4158}
\email{h.jia.cam@gmail.com}
\affiliation{%
  \institution{University of Melbourne}
  \city{Melbourne}
  \state{VIC}
  \country{Australia}
}

\author{Flora D. Salim}
\orcid{0000-0002-1237-1664}
 \email{flora.salim@unsw.edu.au}
\affiliation{%
 \institution{University of New South Wales}
 \streetaddress{Kingsford}
 \city{Sydney}
 \state{NSW}
 \country{Australia}}

\author{Wen Hu}
\orcid{0000-0002-4076-1811}
 \email{wen.hu@unsw.edu.au}
\affiliation{%
 \institution{University of New South Wales}
 \streetaddress{Kingsford}
 \city{Sydney}
 \state{NSW}
 \country{Australia}}

\author{Aaron J. Quigley}
\orcid{0000-0002-5274-6889}
\email{aquigley@acm.org}
\affiliation{%
 \institution{CSIRO's Data61}
 \streetaddress{Kingsford}
 \city{Sydney}
 \state{NSW}
 \country{Australia}}

\renewcommand{\shortauthors}{Yongquan Hu et al.}

\begin{abstract}
Integrating physiological signals such as electroencephalogram (EEG), with other data such as interview audio, may offer valuable multimodal insights into psychological states or neurological disorders. Recent advancements with Large Language Models (LLMs) position them as prospective ``health agents'' for mental health assessment. However, current research predominantly focus on single data modalities, presenting an opportunity to advance understanding through multimodal data. Our study aims to advance this approach by investigating multimodal data using LLMs for mental health assessment, specifically through zero-shot and few-shot prompting. Three datasets are adopted for depression and emotion classifications incorporating EEG, facial expressions, and audio (text). The results indicate that multimodal information confers substantial advantages over single modality approaches in mental health assessment. Notably, integrating EEG alongside commonly used LLM modalities such as audio and images demonstrates promising potential. Moreover, our findings reveal that 1-shot learning offers greater benefits compared to zero-shot learning methods. 
\end{abstract}



\maketitle

\section{Introduction}

Mental health, as defined by the World Health Organization (WHO), is a state of well-being where individuals can realise their potential, handle normal life stresses, work productively, and contribute to their communities \cite{world2022world}. Mental health issues are increasingly impacting the global economy \cite{gao2023critiquing}, with conditions such as depression and anxiety estimated to cost trillions of dollars in lost productivity annually \cite{chisholm2016scaling}. 

The accurate measurement and classification of such health conditions requires psychological evaluation which can include the recording of various indicators. Commonly, many physiological signals, such as Electroencephalogram (EEG) \cite{gore2019surveying}, Heart Rate Variability (HRV) \cite{ha2015wearable}, and Electrodermal Activity (EDA) \cite{greco2016advances}, are integral for mental health assessments due to their reliability and difficulty to mask, ensuring more accurate identification \cite{wang2023detecting,hu2023smartrecorder}. These signals are readily captured by widely available sensors \cite{spathis2019passive,dang2023human,arshad2003exploiting}. 

In addition to capturing data, advancements in Artificial Intelligence (AI) technology have led researchers to develop various algorithms (e.g., machine learning) for the timely and accurately detection \cite{abd2020application}, modeling \cite{xu2023globem} and inference \cite{meegahapola2023generalization} of health conditions based on physiological signals.
Recently, the capabilities of Large-scale Language Models (LLMs) have introduced a new paradigm for prediction and assessment in mental health \cite{lamichhane2023evaluation,wu2024mindshift,xu2024mental}. LLMs offer several advantages, including enhanced multimodal data processing for improved assessment accuracy \cite{li2024omniactions}, interactive communication methods like human-in-the-loop to create more configurable health agents \cite{cabrera2023ethical}, and the potential for fine-tuning domain-specific purpose based on general models to reduce costs \cite{xu2024mental,susnjak2024automating}. However, most work using LLMs to detect mental health focuses on tasks of single modality data such as Mental-LLM \cite{xu2024mental} and EEG-GPT \cite{kim2024eeg}, and the exploration of LLMs in evaluating multimodal sensing data for mental health remains limited. 
Moreover, existing multimodal LLMs have been developed primarily using audio and video modalities. They may lack the capabilities in handling other types of data, such as EEG and other physiological signals which play a crucial role \cite{gore2019surveying} in mental health assessment. Among various physiological signals, EEG is particularly valuable, providing high-frequency data that accurately assesses conditions such as depression, mood, and stress levels \cite{cai2022multi}. Therefore, understanding how these LLMs process EEG data and how to effectively combine EEG with existing modalities remains an open question.

This paper introduces MultiEEG-GPT, a method for assessing mental health using multimodalities, especially with EEG, i.e., EEG and facial expression or audio. The latest GPT-4o API\footnote{https://openai.com/index/hello-GPT-4o/, accessed on June 11, 2024} is adopted for processing multimodalities to recognize the health conditions. Unlike its predecessors such as GPT-4 and GPT-4v, which require separate interface calls, GPT-4o integrates multimodal data processing into a single interface, enhancing the development of this method \cite{OpenAI-GPT-4o}.
This work aims to understand the capabilities of multimodal LLMs in categorising various mental health conditions. This work seeks to compare their ability to model different modalities and EEG and design optimal prompt engineering to facilitate reliable prediction.

The contributions of this paper include: i) the prompt engineering design using both zero-shot and few-shot approaches to examine the predictive capability of MultiEEG-GPT using multimodalities in recognizing different health conditions; ii) experiments across three different databases to validate the effectiveness of MultiEEG-GPT. iii) an in-depth analysis to understand how multimodalities enhance health condition predictions compared to single modalities.
We aim to open up further developments, such as health-supportive social robots~\cite{lai2023psy,hu2023microcam,cabrera2023ethical}, within the context of ubiquitous computing, human-computer interaction, and affective computing.

\section{Related Work} \label{related work}

EEG-based physiological signal analysis has long been essential for monitoring mental health, evolving alongside AI advancements. Initially focused on traditional machine algorithms like k-nearest Neighbor (k-NN) and Support Vector Machine (SVM) for EEG data, Hou et al. demonstrated the potential of EEG for stress level recognition, with the accuracy of 67.07\% \cite{hou2015eeg}. Later, the field has shifted towards integrating deep learning and multimodal data. Zhongjie et al. developed a fusion algorithm levering deep neural networks that combines Convolutional Neural Networks (CNNs) and Bidirectional Long Short-Term Memory (BiLSTM) networks for emotion classification, markedly demonstrating the impressive accuracy in valence and arousal classifications to 93.20±2.55\% and 93.18±2.71\%, respectively \cite{li2021multi}. 

Recently, the advent of general LLMs capable of processing multimodal data has further pivoted the focus towards using LLMs for evaluating mental health data, anticipating their role as future evaluation agents. For example, Xuhai et al. tested various LLMs, including GPT-3.5 and GPT-4, across multiple datasets using methods like zero-shot and few-shot prompting \cite{xu2024mental}. Jonathan et al. introduced EEG-GPT, using GPT models to classify and interpret EEG data \cite{kim2024eeg}. However, these studies still focus on single modality, such as text or EEG. Given various modalities can provide rich and complementary information to infer health conditions, it is proposed to consider different modalities in the automatic systems as well, especially with EEG in many mental health applications. However, research on LLMs for multimodal data with EEG is still limited for mental health prediction. Our proposed MultiEEG-GPT pioneers the work in examining multimodal data including EEG to infer health conditions, aiming to bridge this gap by enhancing the processing of multimodal signals, with a particular focus on EEG data.


\section{Methodology}\label{methodology}
\subsection{Dataset Selection}
Various mental health dataset existed, of which numerous contained EEG modality. Applying the criteria that the dataset need to contain at least EEG modality, we selected 3 most commonly used datasets: 
(1) MODMA \cite{cai2022multi} was developed by Hanshu et al., and this multimodal dataset is designed for analyzing depression disorders and includes oral records (audio) of both patients and controls, and EEG data (convertible to images) from these groups. This dataset has binary labels of whether the participant was diagonsed with Major Depressive Disorder (MDD). 
(2) PME4 \cite{chen2022emotion} 
is a multimodal emotion dataset featuring four modalities: audio, video (not publicly available), EEG, and electromyography (EMG) \cite{chen2022emotion}. It was collected from 11 acting students (five female and six male) who provided informed consent. This dataset focuses on identifying seven emotions: anger, fear, disgust, sadness, happiness, surprise, and a neutral state; 
(3) LUMED-2 \cite{cimtay2020cross} was collected by Loughborough University and Hacettepe University, and it was designed to analyze facial expression, EEG, and galvanic skin response (GSR) data to recognize and classify three categories of human emotions (neutral, happy, sad) under various stimuli
, advancing the understanding in affective computing.

For MODMA and PME4, we used audio and EEG modalities, while for LUMED-2, we used facial expression and EEG modalities. We chose audio and facial expression features because they were the among the most prevalent modalities in mental health analysis \cite{su2020deep,low2020automated}. Besides, the focus of this paper was to explore the possibility of GPT to analyze multimodal data, particularly with the important EEG modality \cite{hickey2021smart}. Thus, we did not include the physiological modalities (e.g., GSR, Resp).

\subsection{Prompt Design}
For our MultiEEG-GPT method, we use prompt engineering strategies (including zero-shot prompting and few-shot prompting) for prediction tasks on multiple datasets. These prompts are model-agnostic, and we present the details of language models and settings employed for our experiment in the next section.

For the prompting strategies, we built upon the design in \cite{xu2024mental} and \cite{xue2023promptcast}. We have designed the prompt to account for different modalities and incorporated flexibility in altering the number of modalities for evaluation. Additionally, we have verified and compared few-shot and zero-shot prompts for evaluation.


\textbf{Zero-shot prompting.} As shown in Table~\ref{tab:prompt_strategy}, the zero-shot prompting strategy consists of a role-play prompt, a specific task description, and an additional rule to avoid unnecessary output and restricted models to focus on the current task. The role-play prompt aims to inform the LLMs of the general task, while the specific task description provides the information for different modalities. Such description also provides the flexibility in adding or deleting modalities. Therefore, the final prompt for the model consisted of: \{role-play prompt\} + \{task specification\} + \{rules\}.

\textbf{Few-shot prompting. }The few-shot prompt added the few-shot samples after the same zero-shot prompt template. Specifically, we include the task-specific prompt followed the zero-shot prompt, but providing the correct class labels instead of offering different candidate class labels for prediction, which is similar to Xuhai et al's setting \cite{xu2024mental}.

\begin{table}[ht!]
\centering
\caption{The zero-shot and few-shot prompting strategies. <MOD1>, <MOD2> and <MOD3> as placeholders denote three different modalities. XXX is the description of collection and visualization process. <SYM> as a placeholder denotes the symptom to be diagnosed. For example, for depression analysis <SYM> should be replaced with depression. The example is for mental health diagonsis with three classes. The label description ``0 denotes XXX'' of the classes could be added or removed to accommodate for more or less classes. }
\begin{tabularx}{0.5\textwidth}{c|X}
\toprule
Role-play prompt & Imagine you are a mental health expert expert at analyzing the emotion and mental health status. \\
\hline
Task specification & The below is <MOD1>, <MOD2> and <MOD3> data. <MOD1> data is collected through XXX and visualized in XXX form. <MOD2> data is collected through XXX and visualized in XXX form. <MOD3> data is collected through XXX and visualized in XXX form. Analyze the <SYM> status of the person. 0 denotes XXX, 1 denotes XXX, 2 denotes XXX. \\ 
\hline
Rules & [Rule]: Do not output other text. \\
\bottomrule
\end{tabularx}
\label{tab:prompt_strategy}
\end{table}


\section{Experiment}\label{experiment}

\begin{table*}[h]
      \caption{Ablation experiment on 3 different multimodal data (EEG image, facial expression, audio). The line with no EEG image, facial expression, audio was determined through majority voting. For few-shot prompting, we chose M=1, which meant we added one few-shot sample in the prompt.}
    \begin{tabular}{@{}lllllll@{}}
        \toprule
        \multicolumn{6}{c}{Prediction Accuracy (\%)}\\ \midrule
        Strategy & EEG & Facial Expression & Audio & MODMA & PME4 & LUMED-2  \\ \midrule
        \multirow{6}{*}{Zero-shot Prompting} & $\times$ & $\times$ & $\times$ & ${50.0_{\pm 0.00}}$ & ${14.28_{\pm 0.00}}$ & ${33.33_{\pm 0.00}}$ \\
        & \checkmark & $\times$ & $\times$ & ${53.79_{\pm 2.46}}$ &  ${21.05_{\pm 1.71}}$  & ${34.61_{\pm 1.28}}$  \\ 
        & $\times$ & \checkmark & $\times$ & -- &  --  & ${38.46_{\pm 1.54}}$     \\ 
        & $\times$ & $\times$ & \checkmark & ${69.35_{\pm 2.53}}$ &  ${15.38_{\pm1.42}}$  & -- \\ 
        \cmidrule{2-7} & \checkmark & \checkmark & $\times$ & -- &  --  & \pmb{${46.13_{\pm 2.42}}$}   \\ 
        & \checkmark & $\times$ & \checkmark & \pmb{${73.54_{\pm 2.03}}$} &  \pmb{${28.57_{\pm 2.41}}$}  & --   \\ \midrule
        \multirow{6}{*}{Few-shot Prompting (M=1)} & $\times$ & $\times$ & $\times$ & ${50.0_{\pm 0.00}}$ & ${14.28_{\pm 0.00}}$ & ${33.33_{\pm 0.00}}$ \\
        & \checkmark & $\times$ & $\times$ & ${62.71_{\pm 3.23}}$ &  ${26.00_{\pm 1.78}}$  & ${36.37_{\pm 1.62}}$  \\ 
        & $\times$ & \checkmark & $\times$ & -- &  --  & ${43.64_{\pm 1.85}}$     \\ 
        & $\times$ & $\times$ & \checkmark & ${69.92_{\pm 1.53}}$ &  ${19.13_{\pm1.29}}$  & -- \\ 
        \cmidrule{2-7} & \checkmark & \checkmark & $\times$ & -- &  --  & \pmb{${52.73_{\pm 2.16}}$}   \\ 
        & \checkmark & $\times$ & \checkmark & \pmb{${79.00_{\pm 1.59}}$} &  \pmb{${37.00_{\pm 2.30}}$}  & --   \\
        \bottomrule
    \end{tabular}
  ~\label{tab:ablation_exp}
\end{table*}

\subsection{Settings}

\subsubsection{Dataset Settings}

\begin{figure*}[h]
    \includegraphics[width=0.9\textwidth]{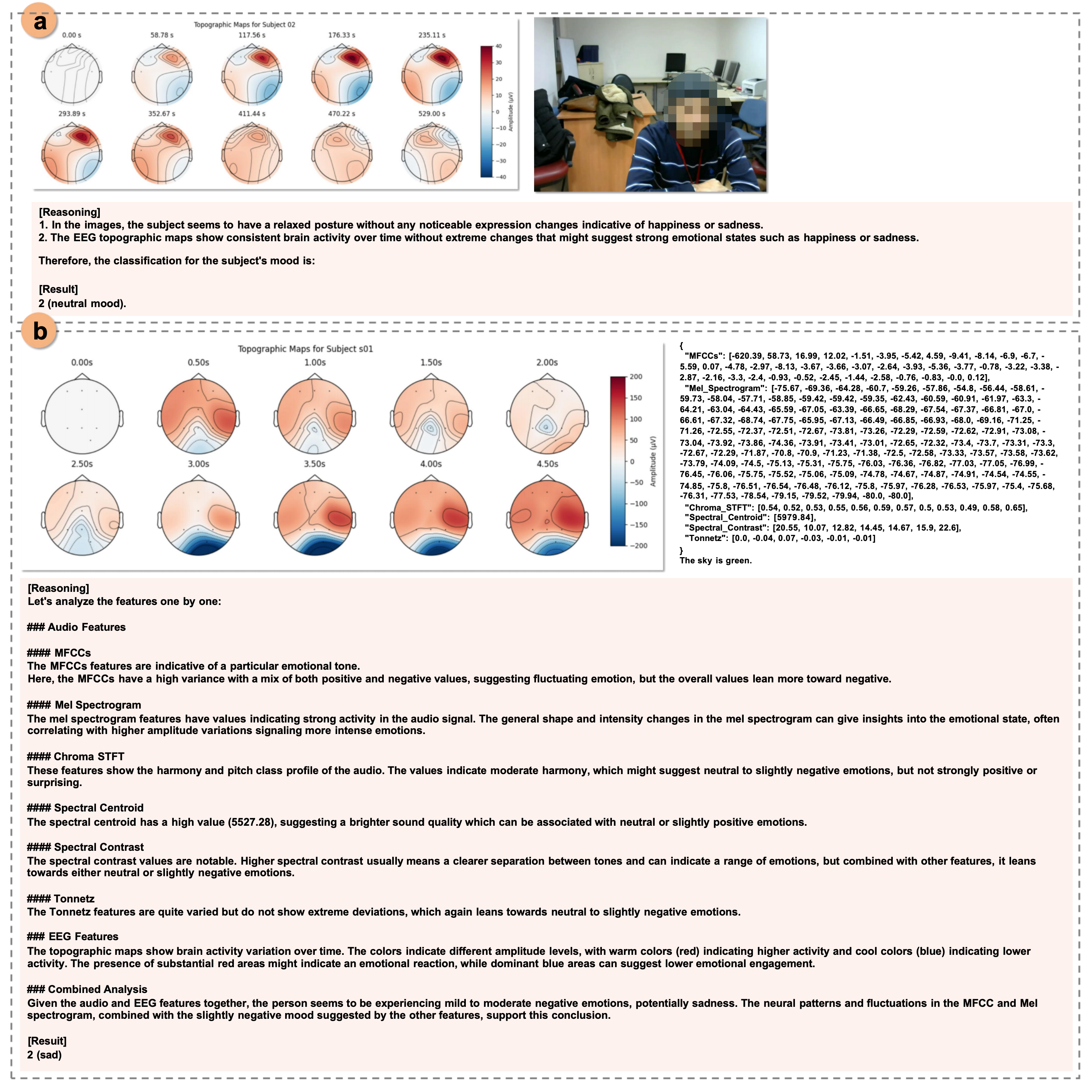}
    \caption{Case analysis for LUMED-2 and PME4 datasets (the person's face has been blurred for ethical reasons). Figure (a) illustrates one subject's input EEG topology map and his facial expression, as well as the prediction result and the text explanation from LUMED-2 dataset. Figure (b) illustrates one subject's input EEG topology map, audio features, input audio transcription ``The sky is green.'', as well as the prediction result and the explanation, from PME4 dataset. In both cases, the model makes the accurate predictions when processing both modalities.}
    \label{fig:case_study}
\end{figure*}

As all the datasets used standard 10-20 electrode layout, we set the electrodes following this layout. MNE library is used for processing EEG signal. 
We processed the datasets using the raw data instead of their pre-processed data (e.g., PME4) because the pre-processed data only contained features instead of the original signals, which were infeasible for plotting topology map. We used a bandpass filter (low-frequency cutoff 0.1Hz, high-frequency cutoff 45Hz, Hamming Window) \cite{pieper2021working} with firwin window design. Afterward, the filtered data were re-referenced to an average reference \cite{pieper2021working}. Since the elicitation presented with different length for different datasets, we chose 530s, 5s and 1.65-4.15s for LUMED-2, PME4 and MODMA datasets respectively, to account for randomly set elicitation time , with 10 equidistant sampled timestamps to create topology maps. For the facial expression, we chose the middle frame of the video (e.g., if the video's length is 10s, we chose the frame at exactly the 5s timestamp) or the image. For the audio, because GPT-4o \footnote{https://community.openai.com/t/when-the-new-voice-model-for-chatgpt-4o-will-be-released/789928, accessed by Jun 16th, 2024} have not yet released the audio input support, we used both the audio features and the text as inputs. For the audio features, we used librosa library to extract the features from the audio and represent these features in text format (which is similar to EEG-GPT's representation \cite{kim2024eeg}), which includes MFCCs, Mel Spectrogram, Chroma STFT, etc. For the text, we transcribed the audio using automatic speech recognition (ASR) systems. We chose the open-sourced vosk library \footnote{https://alphacephei.com/vosk/, accessed by Jun 16th, 2024} with vosk-model-cn-0.15 (Chinese version) or vosk-model-en-0.22 (English version) according to the need. These models were the largest and most advanced ASR systems in the vosk library, which ensured the accuracy of recognition and was used in health care tasks \cite{grasse2021speech,pereira2022web}.

\subsubsection{Model Settings}
For all datasets and all tasks, we transformed the tasks into multi-class classification problems as in previous work \cite{xu2024mental,kim2024eeg}. For MODMA, the binary class labels are 'MDD' or 'healthy'. For PME4, we followed the labels in the original datasets to classify the emotion into seven classes: anger, fear, disgust, sadness, happiness, surprise and neutral. For LUMED-2, we set the 3-class labels as in the original paper, which included neutral, happy ans sadness. 


Previous work showed that GPT-4 generally performed better than GPT-3.5 \cite{xu2024mental}. Given that GPT-4o is the most recent series of GPT-4 that naturally supports multimodal capabilities, we adopted GPT-4o as the tested LLMs. Specifically, we used ``GPT-4o-2024-05-13''\footnote{https://openai.com/index/hello-GPT-4o/, accessed by 11st June, 2024} as the targeted model through OpenAI Azure's API\footnote{https://azure.microsoft.com/en-us/products/ai-services/openai-service, accessed by 11st June, 2024}. For the few-shot experiment, we tested the 1-shot learning scenario to examine the capability of multi-model LLMs with limited information provided. In each repeated trial, we randomly selected one sample from the corresponding dataset to act as the 1-shot sample. This strategy mitigates the bias of selecting samples. For all zero-shot and few-shot experiments, we tested across each dataset (for the few-shot experiment, we excluded that selected sample) for 5 times and reported the average accuracy and the standard deviation.

We use the image updating module of GPT-4o. However, we use no other any additional techniques (e.g., Chain-of-Thoughts \cite{wei2022chain}) to serve as a preliminary study in understanding how multimodal LLMs process multimodal information. This approach ensures the results reflect teh basic capability of the models, which was also consistent with previous work \cite{xu2024mental,kim2024eeg}. 

\subsection{Results and Discussions}

     


\subsubsection{Multimodal analysis.}
We showed two examples of zero-shot cases using LUMED-2 and PME4 dataset in Figure~\ref{fig:case_study}.
The first person in the LUMED-2 video is in neutral mood. The MultiEEG-GPT aims to recognize the participant's mental state through the facial expression and the EEG topology map. As seen in Figure~\ref{fig:case_study}(a), MultiEEG-GPT first processed the image, and then analyzed the EEG topology map. It subsequently aggregated the results from image and EEG jointly, and predicted the participant's emotion state as neural. 

For Figure~\ref{fig:case_study} (b), the participant is in a sad mood. The MultiEEG-GPT first analyzed the person's audio features, and then analyzed the EEG features in the topology map through the color of the map. It finally combined different features and predicted that the participant is in a sad mood. These cases showed the capability of MultiEEG-GPT to (1) analyze each modality separately, (2) aggregated the outputs based on different modalities jointly. It is also evident that a single modality is not adequate to identify the mood correctly. For example, MultiEEG-GPT identified the status of Figure~\ref{fig:case_study} (b) as ``an emotional reaction''. However, it did not accurately state that the participant is sad from the EEG features. By combining the audio features and the EEG features, MUltiEEG-GPT achieved the accurate prediction.

\subsubsection{Performance of MultiEEG-GPT}
Table \ref{tab:ablation_exp} presents the zero-shot and few-shot prompting performance for all three databases. The modalities used for MultiEEG-GPT depend on their availability in the datasets. For zero-shot prompting, our proposed model, utilizing both modalities—either EEG + facial expression or EEG + audio—achieved the best performance compared to other models using a single modality. The proposed model demonstrated relative improvements of 4.19\%, 7.52\%, 7.67\% over the best single-modality performance for the three databases, respectively. This also highlighted the importance of including EEG data in addition to the commonly used modalities in LLMs, such as audio and video. It should be noted that the cases with all modalities removed (the first line) used majority voting similar to Xuhai et al.'s setting \cite{xu2024mental}, serving as the baseline for model performance.

For the few-shot prompting, we observed a similar trend, with multimodal models outperforming single-modality models. Additionally, the 1-shot prompting achieved higher performance than zero-shot prompting, with relative improvements of 5.45\%, 8.43\%, 6.60\% over zero-shot prompting for the MODMA, PME4 and LUMED-2 databases, respectively. This suggests that additional examples enhance recognition, consistent with previous findings \cite{s21144764,liu2024multi}. The extra example likely serves as a benchmark for feature comparison, allowing LLMs to assess the users' mental health status more effectively by comparing features of the few-shot and test samples. The results indicate the general benefit of an additional example, as no specific sample was intentionally selected in the 1-shot prompting setting. In summary, LLMs leveraging multimodalities including EEG could significantly benefit depression and emotion recognition.



\section{Conclusion and Future Work}

This paper proposes MultiEEG-GPT to explore multimodal data, specifically with EEG, for mental health recognition. We have designed zero-shot and few-shot prompting strategies to enhance prediction accuracy, leveraging the most recent GPT-4o as the LLM base model. Three datasets, including MODMA, PME4, and LUMED-2, were adopted for evaluation. Our study showed that predictions using multimodal data significantly outperform those using single-modal data. While the current prediction accuracy approaches that of traditional machine learning methods even without tuning the LLMs, there is significant potential for improvement with strategies such as instruction fine-tuning or multi-strategy hierarchical prediction in future research for mental health leveraging multimodal LLMs.

Moreover, the use of LLMs as health agents raises important ethical considerations. LLMs may exhibit value alignment problems, leading to racial and gender disparities \cite{zack2024assessing} or producing outcomes misaligned with health assessment standards \cite{indran2024twelve}. LLMs also pose privacy risks \cite{peris2023privacy,brown2022does} due to data memorization and extraction \cite{carlini2021extracting}. Fine-tuning with mental health data can lead to data leakage. These issues necessitate careful attention to ensure ethical compliance and accuracy. For example, input data should be anonymized beforehand \cite{staab2024large}, and un-learning and alignment should be integrated to the training process to protect privacy and avoid harm \cite{kirk2024benefits}.

\bibliographystyle{ACM-Reference-Format}
\bibliography{sample}


\begin{thebibliography}{46}


\ifx \showCODEN    \undefined \def \showCODEN     #1{\unskip}     \fi
\ifx \showDOI      \undefined \def \showDOI       #1{#1}\fi
\ifx \showISBNx    \undefined \def \showISBNx     #1{\unskip}     \fi
\ifx \showISBNxiii \undefined \def \showISBNxiii  #1{\unskip}     \fi
\ifx \showISSN     \undefined \def \showISSN      #1{\unskip}     \fi
\ifx \showLCCN     \undefined \def \showLCCN      #1{\unskip}     \fi
\ifx \shownote     \undefined \def \shownote      #1{#1}          \fi
\ifx \showarticletitle \undefined \def \showarticletitle #1{#1}   \fi
\ifx \showURL      \undefined \def \showURL       {\relax}        \fi
\providecommand\bibfield[2]{#2}
\providecommand\bibinfo[2]{#2}
\providecommand\natexlab[1]{#1}
\providecommand\showeprint[2][]{arXiv:#2}

\bibitem[Abd~Rahman et~al\mbox{.}(2020)]%
        {abd2020application}
\bibfield{author}{\bibinfo{person}{Rohizah Abd~Rahman}, \bibinfo{person}{Khairuddin Omar}, \bibinfo{person}{Shahrul Azman~Mohd Noah}, \bibinfo{person}{Mohd Shahrul Nizam~Mohd Danuri}, {and} \bibinfo{person}{Mohammed~Ali Al-Garadi}.} \bibinfo{year}{2020}\natexlab{}.
\newblock \showarticletitle{Application of machine learning methods in mental health detection: a systematic review}.
\newblock \bibinfo{journal}{\emph{Ieee Access}}  \bibinfo{volume}{8} (\bibinfo{year}{2020}), \bibinfo{pages}{183952--183964}.
\newblock


\bibitem[Arshad et~al\mbox{.}(2003)]%
        {arshad2003exploiting}
\bibfield{author}{\bibinfo{person}{Usman Arshad}, \bibinfo{person}{Cecilia Mascolo}, {and} \bibinfo{person}{Marcus Mellor}.} \bibinfo{year}{2003}\natexlab{}.
\newblock \showarticletitle{Exploiting mobile computing in health-care}. In \bibinfo{booktitle}{\emph{Proceedings of demo session of the 3rd international workshop on smart appliances, ICDCS03}}. Citeseer.
\newblock


\bibitem[Brown et~al\mbox{.}(2022)]%
        {brown2022does}
\bibfield{author}{\bibinfo{person}{Hannah Brown}, \bibinfo{person}{Katherine Lee}, \bibinfo{person}{Fatemehsadat Mireshghallah}, \bibinfo{person}{Reza Shokri}, {and} \bibinfo{person}{Florian Tram{\`e}r}.} \bibinfo{year}{2022}\natexlab{}.
\newblock \showarticletitle{What does it mean for a language model to preserve privacy?}. In \bibinfo{booktitle}{\emph{Proceedings of the 2022 ACM conference on fairness, accountability, and transparency}}. \bibinfo{pages}{2280--2292}.
\newblock


\bibitem[Cabrera et~al\mbox{.}(2023)]%
        {cabrera2023ethical}
\bibfield{author}{\bibinfo{person}{Johana Cabrera}, \bibinfo{person}{M~Soledad Loyola}, \bibinfo{person}{Irene Maga{\~n}a}, {and} \bibinfo{person}{Rodrigo Rojas}.} \bibinfo{year}{2023}\natexlab{}.
\newblock \showarticletitle{Ethical dilemmas, mental health, artificial intelligence, and llm-based chatbots}. In \bibinfo{booktitle}{\emph{International Work-Conference on Bioinformatics and Biomedical Engineering}}. Springer, \bibinfo{pages}{313--326}.
\newblock


\bibitem[Cai et~al\mbox{.}(2022)]%
        {cai2022multi}
\bibfield{author}{\bibinfo{person}{Hanshu Cai}, \bibinfo{person}{Zhenqin Yuan}, \bibinfo{person}{Yiwen Gao}, \bibinfo{person}{Shuting Sun}, \bibinfo{person}{Na Li}, \bibinfo{person}{Fuze Tian}, \bibinfo{person}{Han Xiao}, \bibinfo{person}{Jianxiu Li}, \bibinfo{person}{Zhengwu Yang}, \bibinfo{person}{Xiaowei Li}, {et~al\mbox{.}}} \bibinfo{year}{2022}\natexlab{}.
\newblock \showarticletitle{A multi-modal open dataset for mental-disorder analysis}.
\newblock \bibinfo{journal}{\emph{Scientific Data}} \bibinfo{volume}{9}, \bibinfo{number}{1} (\bibinfo{year}{2022}), \bibinfo{pages}{178}.
\newblock


\bibitem[Carlini et~al\mbox{.}(2021)]%
        {carlini2021extracting}
\bibfield{author}{\bibinfo{person}{Nicholas Carlini}, \bibinfo{person}{Florian Tramer}, \bibinfo{person}{Eric Wallace}, \bibinfo{person}{Matthew Jagielski}, \bibinfo{person}{Ariel Herbert-Voss}, \bibinfo{person}{Katherine Lee}, \bibinfo{person}{Adam Roberts}, \bibinfo{person}{Tom Brown}, \bibinfo{person}{Dawn Song}, \bibinfo{person}{Ulfar Erlingsson}, {et~al\mbox{.}}} \bibinfo{year}{2021}\natexlab{}.
\newblock \showarticletitle{Extracting training data from large language models}. In \bibinfo{booktitle}{\emph{30th USENIX Security Symposium (USENIX Security 21)}}. \bibinfo{pages}{2633--2650}.
\newblock


\bibitem[Chen et~al\mbox{.}(2022)]%
        {chen2022emotion}
\bibfield{author}{\bibinfo{person}{Jin Chen}, \bibinfo{person}{Tony Ro}, {and} \bibinfo{person}{Zhigang Zhu}.} \bibinfo{year}{2022}\natexlab{}.
\newblock \showarticletitle{Emotion recognition with audio, video, EEG, and EMG: a dataset and baseline approaches}.
\newblock \bibinfo{journal}{\emph{IEEE Access}}  \bibinfo{volume}{10} (\bibinfo{year}{2022}), \bibinfo{pages}{13229--13242}.
\newblock


\bibitem[Chisholm et~al\mbox{.}(2016)]%
        {chisholm2016scaling}
\bibfield{author}{\bibinfo{person}{Dan Chisholm}, \bibinfo{person}{Kim Sweeny}, \bibinfo{person}{Peter Sheehan}, \bibinfo{person}{Bruce Rasmussen}, \bibinfo{person}{Filip Smit}, \bibinfo{person}{Pim Cuijpers}, {and} \bibinfo{person}{Shekhar Saxena}.} \bibinfo{year}{2016}\natexlab{}.
\newblock \showarticletitle{Scaling-up treatment of depression and anxiety: a global return on investment analysis}.
\newblock \bibinfo{journal}{\emph{The Lancet Psychiatry}} \bibinfo{volume}{3}, \bibinfo{number}{5} (\bibinfo{year}{2016}), \bibinfo{pages}{415--424}.
\newblock


\bibitem[Cimtay et~al\mbox{.}(2020)]%
        {cimtay2020cross}
\bibfield{author}{\bibinfo{person}{Yucel Cimtay}, \bibinfo{person}{Erhan Ekmekcioglu}, {and} \bibinfo{person}{Seyma Caglar-Ozhan}.} \bibinfo{year}{2020}\natexlab{}.
\newblock \showarticletitle{Cross-subject multimodal emotion recognition based on hybrid fusion}.
\newblock \bibinfo{journal}{\emph{IEEE Access}}  \bibinfo{volume}{8} (\bibinfo{year}{2020}), \bibinfo{pages}{168865--168878}.
\newblock


\bibitem[Dang et~al\mbox{.}(2023)]%
        {dang2023human}
\bibfield{author}{\bibinfo{person}{Ting Dang}, \bibinfo{person}{Dimitris Spathis}, \bibinfo{person}{Abhirup Ghosh}, {and} \bibinfo{person}{Cecilia Mascolo}.} \bibinfo{year}{2023}\natexlab{}.
\newblock \showarticletitle{Human-centred artificial intelligence for mobile health sensing: challenges and opportunities}.
\newblock \bibinfo{journal}{\emph{Royal Society Open Science}} \bibinfo{volume}{10}, \bibinfo{number}{11} (\bibinfo{year}{2023}), \bibinfo{pages}{230806}.
\newblock


\bibitem[Gao et~al\mbox{.}(2023)]%
        {gao2023critiquing}
\bibfield{author}{\bibinfo{person}{Nan Gao}, \bibinfo{person}{Soundariya Ananthan}, \bibinfo{person}{Chun Yu}, \bibinfo{person}{Yuntao Wang}, {and} \bibinfo{person}{Flora~D Salim}.} \bibinfo{year}{2023}\natexlab{}.
\newblock \showarticletitle{Critiquing Self-report Practices for Human Mental and Wellbeing Computing at Ubicomp}.
\newblock \bibinfo{journal}{\emph{arXiv preprint arXiv:2311.15496}} (\bibinfo{year}{2023}).
\newblock


\bibitem[Gore and Rathi(2019)]%
        {gore2019surveying}
\bibfield{author}{\bibinfo{person}{Ela Gore} {and} \bibinfo{person}{Sheetal Rathi}.} \bibinfo{year}{2019}\natexlab{}.
\newblock \showarticletitle{Surveying machine learning algorithms on eeg signals data for mental health assessment}. In \bibinfo{booktitle}{\emph{2019 IEEE Pune Section International Conference (PuneCon)}}. IEEE, \bibinfo{pages}{1--6}.
\newblock


\bibitem[Grasse et~al\mbox{.}(2021)]%
        {grasse2021speech}
\bibfield{author}{\bibinfo{person}{Lukas Grasse}, \bibinfo{person}{Sylvain~J Boutros}, {and} \bibinfo{person}{Matthew~S Tata}.} \bibinfo{year}{2021}\natexlab{}.
\newblock \showarticletitle{Speech interaction to control a hands-free delivery robot for high-risk health care scenarios}.
\newblock \bibinfo{journal}{\emph{Frontiers in Robotics and AI}}  \bibinfo{volume}{8} (\bibinfo{year}{2021}), \bibinfo{pages}{612750}.
\newblock


\bibitem[Greco et~al\mbox{.}(2016)]%
        {greco2016advances}
\bibfield{author}{\bibinfo{person}{Alberto Greco}, \bibinfo{person}{Gaetano Valenza}, {and} \bibinfo{person}{Enzo~Pasquale Scilingo}.} \bibinfo{year}{2016}\natexlab{}.
\newblock \bibinfo{booktitle}{\emph{Advances in Electrodermal activity processing with applications for mental health}}.
\newblock \bibinfo{publisher}{Springer}.
\newblock


\bibitem[Ha et~al\mbox{.}(2015)]%
        {ha2015wearable}
\bibfield{author}{\bibinfo{person}{Unsoo Ha}, \bibinfo{person}{Yongsu Lee}, \bibinfo{person}{Hyunki Kim}, \bibinfo{person}{Taehwan Roh}, \bibinfo{person}{Joonsung Bae}, \bibinfo{person}{Changhyeon Kim}, {and} \bibinfo{person}{Hoi-Jun Yoo}.} \bibinfo{year}{2015}\natexlab{}.
\newblock \showarticletitle{A wearable EEG-HEG-HRV multimodal system with simultaneous monitoring of tES for mental health management}.
\newblock \bibinfo{journal}{\emph{IEEE transactions on biomedical circuits and systems}} \bibinfo{volume}{9}, \bibinfo{number}{6} (\bibinfo{year}{2015}), \bibinfo{pages}{758--766}.
\newblock


\bibitem[Hickey et~al\mbox{.}(2021)]%
        {hickey2021smart}
\bibfield{author}{\bibinfo{person}{Blake~Anthony Hickey}, \bibinfo{person}{Taryn Chalmers}, \bibinfo{person}{Phillip Newton}, \bibinfo{person}{Chin-Teng Lin}, \bibinfo{person}{David Sibbritt}, \bibinfo{person}{Craig~S McLachlan}, \bibinfo{person}{Roderick Clifton-Bligh}, \bibinfo{person}{John Morley}, {and} \bibinfo{person}{Sara Lal}.} \bibinfo{year}{2021}\natexlab{}.
\newblock \showarticletitle{Smart devices and wearable technologies to detect and monitor mental health conditions and stress: A systematic review}.
\newblock \bibinfo{journal}{\emph{Sensors}} \bibinfo{volume}{21}, \bibinfo{number}{10} (\bibinfo{year}{2021}), \bibinfo{pages}{3461}.
\newblock


\bibitem[Hou et~al\mbox{.}(2015)]%
        {hou2015eeg}
\bibfield{author}{\bibinfo{person}{Xiyuan Hou}, \bibinfo{person}{Yisi Liu}, \bibinfo{person}{Olga Sourina}, \bibinfo{person}{Yun Rui~Eileen Tan}, \bibinfo{person}{Lipo Wang}, {and} \bibinfo{person}{Wolfgang Mueller-Wittig}.} \bibinfo{year}{2015}\natexlab{}.
\newblock \showarticletitle{EEG based stress monitoring}. In \bibinfo{booktitle}{\emph{2015 IEEE international conference on systems, man, and cybernetics}}. IEEE, \bibinfo{pages}{3110--3115}.
\newblock


\bibitem[Hu et~al\mbox{.}(2023a)]%
        {hu2023smartrecorder}
\bibfield{author}{\bibinfo{person}{Xiaozhu Hu}, \bibinfo{person}{Yanwen Huang}, \bibinfo{person}{Bo Liu}, \bibinfo{person}{Ruolan Wu}, \bibinfo{person}{Yongquan Hu}, \bibinfo{person}{Aaron~J Quigley}, \bibinfo{person}{Mingming Fan}, \bibinfo{person}{Chun Yu}, {and} \bibinfo{person}{Yuanchun Shi}.} \bibinfo{year}{2023}\natexlab{a}.
\newblock \showarticletitle{SmartRecorder: An IMU-based Video Tutorial Creation by Demonstration System for Smartphone Interaction Tasks}. In \bibinfo{booktitle}{\emph{Proceedings of the 28th International Conference on Intelligent User Interfaces}}. \bibinfo{pages}{278--293}.
\newblock


\bibitem[Hu et~al\mbox{.}(2023b)]%
        {hu2023microcam}
\bibfield{author}{\bibinfo{person}{Yongquan Hu}, \bibinfo{person}{Hui-Shyong Yeo}, \bibinfo{person}{Mingyue Yuan}, \bibinfo{person}{Haoran Fan}, \bibinfo{person}{Don~Samitha Elvitigala}, \bibinfo{person}{Wen Hu}, {and} \bibinfo{person}{Aaron Quigley}.} \bibinfo{year}{2023}\natexlab{b}.
\newblock \showarticletitle{Microcam: Leveraging smartphone microscope camera for context-aware contact surface sensing}.
\newblock \bibinfo{journal}{\emph{Proceedings of the ACM on Interactive, Mobile, Wearable and Ubiquitous Technologies}} \bibinfo{volume}{7}, \bibinfo{number}{3} (\bibinfo{year}{2023}), \bibinfo{pages}{1--28}.
\newblock


\bibitem[Indran et~al\mbox{.}(2024)]%
        {indran2024twelve}
\bibfield{author}{\bibinfo{person}{Inthrani~Raja Indran}, \bibinfo{person}{Priya Paranthaman}, \bibinfo{person}{Neelima Gupta}, {and} \bibinfo{person}{Nurulhuda Mustafa}.} \bibinfo{year}{2024}\natexlab{}.
\newblock \showarticletitle{Twelve tips to leverage AI for efficient and effective medical question generation: a guide for educators using Chat GPT}.
\newblock \bibinfo{journal}{\emph{Medical Teacher}} (\bibinfo{year}{2024}), \bibinfo{pages}{1--6}.
\newblock


\bibitem[Kim et~al\mbox{.}(2024)]%
        {kim2024eeg}
\bibfield{author}{\bibinfo{person}{Jonathan~W Kim}, \bibinfo{person}{Ahmed Alaa}, {and} \bibinfo{person}{Danilo Bernardo}.} \bibinfo{year}{2024}\natexlab{}.
\newblock \showarticletitle{EEG-GPT: Exploring Capabilities of Large Language Models for EEG Classification and Interpretation}.
\newblock \bibinfo{journal}{\emph{arXiv preprint arXiv:2401.18006}} (\bibinfo{year}{2024}).
\newblock


\bibitem[Kirk et~al\mbox{.}(2024)]%
        {kirk2024benefits}
\bibfield{author}{\bibinfo{person}{Hannah~Rose Kirk}, \bibinfo{person}{Bertie Vidgen}, \bibinfo{person}{Paul R{\"o}ttger}, {and} \bibinfo{person}{Scott~A Hale}.} \bibinfo{year}{2024}\natexlab{}.
\newblock \showarticletitle{The benefits, risks and bounds of personalizing the alignment of large language models to individuals}.
\newblock \bibinfo{journal}{\emph{Nature Machine Intelligence}} (\bibinfo{year}{2024}), \bibinfo{pages}{1--10}.
\newblock


\bibitem[Lai et~al\mbox{.}(2023)]%
        {lai2023psy}
\bibfield{author}{\bibinfo{person}{Tin Lai}, \bibinfo{person}{Yukun Shi}, \bibinfo{person}{Zicong Du}, \bibinfo{person}{Jiajie Wu}, \bibinfo{person}{Ken Fu}, \bibinfo{person}{Yichao Dou}, {and} \bibinfo{person}{Ziqi Wang}.} \bibinfo{year}{2023}\natexlab{}.
\newblock \showarticletitle{Psy-llm: Scaling up global mental health psychological services with ai-based large language models}.
\newblock \bibinfo{journal}{\emph{arXiv preprint arXiv:2307.11991}} (\bibinfo{year}{2023}).
\newblock


\bibitem[Lamichhane(2023)]%
        {lamichhane2023evaluation}
\bibfield{author}{\bibinfo{person}{Bishal Lamichhane}.} \bibinfo{year}{2023}\natexlab{}.
\newblock \showarticletitle{Evaluation of chatgpt for nlp-based mental health applications}.
\newblock \bibinfo{journal}{\emph{arXiv preprint arXiv:2303.15727}} (\bibinfo{year}{2023}).
\newblock


\bibitem[Li et~al\mbox{.}(2024)]%
        {li2024omniactions}
\bibfield{author}{\bibinfo{person}{Jiahao~Nick Li}, \bibinfo{person}{Yan Xu}, \bibinfo{person}{Tovi Grossman}, \bibinfo{person}{Stephanie Santosa}, {and} \bibinfo{person}{Michelle Li}.} \bibinfo{year}{2024}\natexlab{}.
\newblock \showarticletitle{OmniActions: Predicting Digital Actions in Response to Real-World Multimodal Sensory Inputs with LLMs}. In \bibinfo{booktitle}{\emph{Proceedings of the CHI Conference on Human Factors in Computing Systems}}. \bibinfo{pages}{1--22}.
\newblock


\bibitem[Li et~al\mbox{.}(2021)]%
        {li2021multi}
\bibfield{author}{\bibinfo{person}{Zhongjie Li}, \bibinfo{person}{Gaoyan Zhang}, \bibinfo{person}{Jianwu Dang}, \bibinfo{person}{Longbiao Wang}, {and} \bibinfo{person}{Jianguo Wei}.} \bibinfo{year}{2021}\natexlab{}.
\newblock \showarticletitle{Multi-modal emotion recognition based on deep learning of EEG and audio signals}. In \bibinfo{booktitle}{\emph{2021 International Joint Conference on Neural Networks (IJCNN)}}. IEEE, \bibinfo{pages}{1--6}.
\newblock


\bibitem[Liu et~al\mbox{.}(2024)]%
        {liu2024multi}
\bibfield{author}{\bibinfo{person}{Liangliang Liu}, \bibinfo{person}{Zhihong Liu}, \bibinfo{person}{Jing Chang}, {and} \bibinfo{person}{Xue Xu}.} \bibinfo{year}{2024}\natexlab{}.
\newblock \showarticletitle{A multi-modal extraction integrated model for neuropsychiatric disorders classification}.
\newblock \bibinfo{journal}{\emph{Pattern Recognition}} (\bibinfo{year}{2024}), \bibinfo{pages}{110646}.
\newblock


\bibitem[Low et~al\mbox{.}(2020)]%
        {low2020automated}
\bibfield{author}{\bibinfo{person}{Daniel~M Low}, \bibinfo{person}{Kate~H Bentley}, {and} \bibinfo{person}{Satrajit~S Ghosh}.} \bibinfo{year}{2020}\natexlab{}.
\newblock \showarticletitle{Automated assessment of psychiatric disorders using speech: A systematic review}.
\newblock \bibinfo{journal}{\emph{Laryngoscope investigative otolaryngology}} \bibinfo{volume}{5}, \bibinfo{number}{1} (\bibinfo{year}{2020}), \bibinfo{pages}{96--116}.
\newblock


\bibitem[Meegahapola et~al\mbox{.}(2023)]%
        {meegahapola2023generalization}
\bibfield{author}{\bibinfo{person}{Lakmal Meegahapola}, \bibinfo{person}{William Droz}, \bibinfo{person}{Peter Kun}, \bibinfo{person}{Amalia De~G{\"o}tzen}, \bibinfo{person}{Chaitanya Nutakki}, \bibinfo{person}{Shyam Diwakar}, \bibinfo{person}{Salvador~Ruiz Correa}, \bibinfo{person}{Donglei Song}, \bibinfo{person}{Hao Xu}, \bibinfo{person}{Miriam Bidoglia}, {et~al\mbox{.}}} \bibinfo{year}{2023}\natexlab{}.
\newblock \showarticletitle{Generalization and personalization of mobile sensing-based mood inference models: an analysis of college students in eight countries}.
\newblock \bibinfo{journal}{\emph{Proceedings of the ACM on interactive, mobile, wearable and ubiquitous technologies}} \bibinfo{volume}{6}, \bibinfo{number}{4} (\bibinfo{year}{2023}), \bibinfo{pages}{1--32}.
\newblock


\bibitem[OpenAI(2024)]%
        {OpenAI-GPT-4o}
\bibfield{author}{\bibinfo{person}{OpenAI}.} \bibinfo{year}{2024}\natexlab{}.
\newblock \bibinfo{title}{Hello GPT-4o}.
\newblock
\newblock
\newblock
\shownote{\url{https://openai.com/index/hello-gpt-4o/} [Accessed:June 2024]}.


\bibitem[Organization et~al\mbox{.}(2022)]%
        {world2022world}
\bibfield{author}{\bibinfo{person}{World~Health Organization} {et~al\mbox{.}}} \bibinfo{year}{2022}\natexlab{}.
\newblock \showarticletitle{World mental health report: Transforming mental health for all}.
\newblock  (\bibinfo{year}{2022}).
\newblock


\bibitem[Pereira et~al\mbox{.}(2022)]%
        {pereira2022web}
\bibfield{author}{\bibinfo{person}{Tiago~F Pereira}, \bibinfo{person}{Arthur Matta}, \bibinfo{person}{Carlos~M Mayea}, \bibinfo{person}{Frederico Pereira}, \bibinfo{person}{Nelson Monroy}, \bibinfo{person}{Jo{\~a}o Jorge}, \bibinfo{person}{Tiago Rosa}, \bibinfo{person}{Carlos~E Salgado}, \bibinfo{person}{Ana Lima}, \bibinfo{person}{Ricardo~J Machado}, {et~al\mbox{.}}} \bibinfo{year}{2022}\natexlab{}.
\newblock \showarticletitle{A web-based Voice Interaction framework proposal for enhancing Information Systems user experience}.
\newblock \bibinfo{journal}{\emph{Procedia Computer Science}}  \bibinfo{volume}{196} (\bibinfo{year}{2022}), \bibinfo{pages}{235--244}.
\newblock


\bibitem[Peris et~al\mbox{.}(2023)]%
        {peris2023privacy}
\bibfield{author}{\bibinfo{person}{Charith Peris}, \bibinfo{person}{Christophe Dupuy}, \bibinfo{person}{Jimit Majmudar}, \bibinfo{person}{Rahil Parikh}, \bibinfo{person}{Sami Smaili}, \bibinfo{person}{Richard Zemel}, {and} \bibinfo{person}{Rahul Gupta}.} \bibinfo{year}{2023}\natexlab{}.
\newblock \showarticletitle{Privacy in the time of language models}. In \bibinfo{booktitle}{\emph{Proceedings of the sixteenth ACM international conference on web search and data mining}}. \bibinfo{pages}{1291--1292}.
\newblock


\bibitem[Pieper et~al\mbox{.}(2021)]%
        {pieper2021working}
\bibfield{author}{\bibinfo{person}{Kerstin Pieper}, \bibinfo{person}{Robert~P Spang}, \bibinfo{person}{Pablo Prietz}, \bibinfo{person}{Sebastian M{\"o}ller}, \bibinfo{person}{Erkki Paajanen}, \bibinfo{person}{Markus Vaalgamaa}, {and} \bibinfo{person}{Jan-Niklas Voigt-Antons}.} \bibinfo{year}{2021}\natexlab{}.
\newblock \showarticletitle{Working with environmental noise and noise-cancelation: a workload assessment with EEG and subjective measures}.
\newblock \bibinfo{journal}{\emph{Frontiers in neuroscience}}  \bibinfo{volume}{15} (\bibinfo{year}{2021}), \bibinfo{pages}{771533}.
\newblock


\bibitem[Spathis et~al\mbox{.}(2019)]%
        {spathis2019passive}
\bibfield{author}{\bibinfo{person}{Dimitris Spathis}, \bibinfo{person}{Sandra Servia-Rodriguez}, \bibinfo{person}{Katayoun Farrahi}, \bibinfo{person}{Cecilia Mascolo}, {and} \bibinfo{person}{Jason Rentfrow}.} \bibinfo{year}{2019}\natexlab{}.
\newblock \showarticletitle{Passive mobile sensing and psychological traits for large scale mood prediction}. In \bibinfo{booktitle}{\emph{Proceedings of the 13th EAI international conference on pervasive computing technologies for healthcare}}. \bibinfo{pages}{272--281}.
\newblock


\bibitem[Staab et~al\mbox{.}(2024)]%
        {staab2024large}
\bibfield{author}{\bibinfo{person}{Robin Staab}, \bibinfo{person}{Mark Vero}, \bibinfo{person}{Mislav Balunovic}, {and} \bibinfo{person}{Martin Vechev}.} \bibinfo{year}{2024}\natexlab{}.
\newblock \showarticletitle{Large Language Models are Anonymizers}. In \bibinfo{booktitle}{\emph{ICLR 2024 Workshop on Reliable and Responsible Foundation Models}}.
\newblock


\bibitem[Su et~al\mbox{.}(2020)]%
        {su2020deep}
\bibfield{author}{\bibinfo{person}{Chang Su}, \bibinfo{person}{Zhenxing Xu}, \bibinfo{person}{Jyotishman Pathak}, {and} \bibinfo{person}{Fei Wang}.} \bibinfo{year}{2020}\natexlab{}.
\newblock \showarticletitle{Deep learning in mental health outcome research: a scoping review}.
\newblock \bibinfo{journal}{\emph{Translational Psychiatry}} \bibinfo{volume}{10}, \bibinfo{number}{1} (\bibinfo{year}{2020}), \bibinfo{pages}{116}.
\newblock


\bibitem[Sun et~al\mbox{.}(2021)]%
        {s21144764}
\bibfield{author}{\bibinfo{person}{Hao Sun}, \bibinfo{person}{Jiaqing Liu}, \bibinfo{person}{Shurong Chai}, \bibinfo{person}{Zhaolin Qiu}, \bibinfo{person}{Lanfen Lin}, \bibinfo{person}{Xinyin Huang}, {and} \bibinfo{person}{Yenwei Chen}.} \bibinfo{year}{2021}\natexlab{}.
\newblock \showarticletitle{Multi-Modal Adaptive Fusion Transformer Network for the Estimation of Depression Level}.
\newblock \bibinfo{journal}{\emph{Sensors}} \bibinfo{volume}{21}, \bibinfo{number}{14} (\bibinfo{year}{2021}).
\newblock
\showISSN{1424-8220}
\urldef\tempurl%
\url{https://doi.org/10.3390/s21144764}
\showDOI{\tempurl}


\bibitem[Susnjak et~al\mbox{.}(2024)]%
        {susnjak2024automating}
\bibfield{author}{\bibinfo{person}{Teo Susnjak}, \bibinfo{person}{Peter Hwang}, \bibinfo{person}{Napoleon~H Reyes}, \bibinfo{person}{Andre~LC Barczak}, \bibinfo{person}{Timothy~R McIntosh}, {and} \bibinfo{person}{Surangika Ranathunga}.} \bibinfo{year}{2024}\natexlab{}.
\newblock \showarticletitle{Automating research synthesis with domain-specific large language model fine-tuning}.
\newblock \bibinfo{journal}{\emph{arXiv preprint arXiv:2404.08680}} (\bibinfo{year}{2024}).
\newblock


\bibitem[Wang et~al\mbox{.}(2023)]%
        {wang2023detecting}
\bibfield{author}{\bibinfo{person}{Zhiyuan Wang}, \bibinfo{person}{Maria~A Larrazabal}, \bibinfo{person}{Mark Rucker}, \bibinfo{person}{Emma~R Toner}, \bibinfo{person}{Katharine~E Daniel}, \bibinfo{person}{Shashwat Kumar}, \bibinfo{person}{Mehdi Boukhechba}, \bibinfo{person}{Bethany~A Teachman}, {and} \bibinfo{person}{Laura~E Barnes}.} \bibinfo{year}{2023}\natexlab{}.
\newblock \showarticletitle{Detecting social contexts from mobile sensing indicators in virtual interactions with socially anxious individuals}.
\newblock \bibinfo{journal}{\emph{Proceedings of the ACM on Interactive, Mobile, Wearable and Ubiquitous Technologies}} \bibinfo{volume}{7}, \bibinfo{number}{3} (\bibinfo{year}{2023}), \bibinfo{pages}{1--26}.
\newblock


\bibitem[Wei et~al\mbox{.}(2022)]%
        {wei2022chain}
\bibfield{author}{\bibinfo{person}{Jason Wei}, \bibinfo{person}{Xuezhi Wang}, \bibinfo{person}{Dale Schuurmans}, \bibinfo{person}{Maarten Bosma}, \bibinfo{person}{Fei Xia}, \bibinfo{person}{Ed Chi}, \bibinfo{person}{Quoc~V Le}, \bibinfo{person}{Denny Zhou}, {et~al\mbox{.}}} \bibinfo{year}{2022}\natexlab{}.
\newblock \showarticletitle{Chain-of-thought prompting elicits reasoning in large language models}.
\newblock \bibinfo{journal}{\emph{Advances in neural information processing systems}}  \bibinfo{volume}{35} (\bibinfo{year}{2022}), \bibinfo{pages}{24824--24837}.
\newblock


\bibitem[Wu et~al\mbox{.}(2024)]%
        {wu2024mindshift}
\bibfield{author}{\bibinfo{person}{Ruolan Wu}, \bibinfo{person}{Chun Yu}, \bibinfo{person}{Xiaole Pan}, \bibinfo{person}{Yujia Liu}, \bibinfo{person}{Ningning Zhang}, \bibinfo{person}{Yue Fu}, \bibinfo{person}{Yuhan Wang}, \bibinfo{person}{Zhi Zheng}, \bibinfo{person}{Li Chen}, \bibinfo{person}{Qiaolei Jiang}, {et~al\mbox{.}}} \bibinfo{year}{2024}\natexlab{}.
\newblock \showarticletitle{MindShift: Leveraging Large Language Models for Mental-States-Based Problematic Smartphone Use Intervention}. In \bibinfo{booktitle}{\emph{Proceedings of the CHI Conference on Human Factors in Computing Systems}}. \bibinfo{pages}{1--24}.
\newblock


\bibitem[Xu et~al\mbox{.}(2023)]%
        {xu2023globem}
\bibfield{author}{\bibinfo{person}{Xuhai Xu}, \bibinfo{person}{Xin Liu}, \bibinfo{person}{Han Zhang}, \bibinfo{person}{Weichen Wang}, \bibinfo{person}{Subigya Nepal}, \bibinfo{person}{Yasaman Sefidgar}, \bibinfo{person}{Woosuk Seo}, \bibinfo{person}{Kevin~S Kuehn}, \bibinfo{person}{Jeremy~F Huckins}, \bibinfo{person}{Margaret~E Morris}, {et~al\mbox{.}}} \bibinfo{year}{2023}\natexlab{}.
\newblock \showarticletitle{GLOBEM: cross-dataset generalization of longitudinal human behavior modeling}.
\newblock \bibinfo{journal}{\emph{Proceedings of the ACM on Interactive, Mobile, Wearable and Ubiquitous Technologies}} \bibinfo{volume}{6}, \bibinfo{number}{4} (\bibinfo{year}{2023}), \bibinfo{pages}{1--34}.
\newblock


\bibitem[Xu et~al\mbox{.}(2024)]%
        {xu2024mental}
\bibfield{author}{\bibinfo{person}{Xuhai Xu}, \bibinfo{person}{Bingsheng Yao}, \bibinfo{person}{Yuanzhe Dong}, \bibinfo{person}{Saadia Gabriel}, \bibinfo{person}{Hong Yu}, \bibinfo{person}{James Hendler}, \bibinfo{person}{Marzyeh Ghassemi}, \bibinfo{person}{Anind~K Dey}, {and} \bibinfo{person}{Dakuo Wang}.} \bibinfo{year}{2024}\natexlab{}.
\newblock \showarticletitle{Mental-llm: Leveraging large language models for mental health prediction via online text data}.
\newblock \bibinfo{journal}{\emph{Proceedings of the ACM on Interactive, Mobile, Wearable and Ubiquitous Technologies}} \bibinfo{volume}{8}, \bibinfo{number}{1} (\bibinfo{year}{2024}), \bibinfo{pages}{1--32}.
\newblock


\bibitem[Xue and Salim(2023)]%
        {xue2023promptcast}
\bibfield{author}{\bibinfo{person}{Hao Xue} {and} \bibinfo{person}{Flora~D Salim}.} \bibinfo{year}{2023}\natexlab{}.
\newblock \showarticletitle{Promptcast: A new prompt-based learning paradigm for time series forecasting}.
\newblock \bibinfo{journal}{\emph{IEEE Transactions on Knowledge and Data Engineering}} (\bibinfo{year}{2023}).
\newblock


\bibitem[Zack et~al\mbox{.}(2024)]%
        {zack2024assessing}
\bibfield{author}{\bibinfo{person}{Travis Zack}, \bibinfo{person}{Eric Lehman}, \bibinfo{person}{Mirac Suzgun}, \bibinfo{person}{Jorge~A Rodriguez}, \bibinfo{person}{Leo~Anthony Celi}, \bibinfo{person}{Judy Gichoya}, \bibinfo{person}{Dan Jurafsky}, \bibinfo{person}{Peter Szolovits}, \bibinfo{person}{David~W Bates}, \bibinfo{person}{Raja-Elie~E Abdulnour}, {et~al\mbox{.}}} \bibinfo{year}{2024}\natexlab{}.
\newblock \showarticletitle{Assessing the potential of GPT-4 to perpetuate racial and gender biases in health care: a model evaluation study}.
\newblock \bibinfo{journal}{\emph{The Lancet Digital Health}} \bibinfo{volume}{6}, \bibinfo{number}{1} (\bibinfo{year}{2024}), \bibinfo{pages}{e12--e22}.
\newblock


\end{thebibliography}


\end{document}